\documentclass[a4paper,fleqn,3p]{elsarticle}
\usepackage{hyperref}
\usepackage{subcaption}
\usepackage{textcomp}
\usepackage{multirow}
\usepackage{multicol}\usepackage{array}
\usepackage{booktabs}
\usepackage{xcolor}

\usepackage[LGRgreek]{mathastext}

\begin{document}

\begin{frontmatter}

\title{$^{222}$Rn content variations at ground and underground conditions}

\address[AR]{Baksan Neutrino Observatory INR RAS, Neitrino 361609, Russia}
\address[BR]{H.M.Berbekov Kabardino-Balkarian State University, Nalchik 360004, Russia}
\address[KR]{V.N.~Karazin Kharkiv National University, Kharkiv, 61022 Ukraine}

\author[AR]{R.A.~Etezov}
\author[AR]{Yu.M.~Gavrilyuk}
\author[AR,BR]{A.M.~Gangapshev}
\author[AR]{A.M.~Gezhaev}
\author[AR]{V.V.~Kazalov}
\author[AR,BR]{A.Kh.~Khokonov}
\author[AR,BR]{V.V.~Kuzminov\corref{cor1}}
\author[KR]{S.I.~Panasenko}
\author[KR]{S.S.~Ratkevich\corref{cor2}}

\cortext[cor1]{{bno$\_$vvk@mail.ru}}
\cortext[cor2]{{ssratkevich@karazin.ua}}

\begin{abstract}
The activity of $^{222}$Rn and its daughter isotopes was measured in the air of several underground laboratories of the Baksan Neutrino Observatory at various distances from the entrance. The measurements were carried out with the help of the cylindrical ionionization air chamber.
We found that the radon content in the ventilated airflow within the measurement accuracy does not depend on the distance traveled along the adit. In addition, we observed that the radon content increases abruptly in those locations where underground gases and water are released. As a result, we review various mechanisms of air enrichment with radon. We also outline our research methodology and present the results of our measurements of radon release from the rocky walls of the underground laboratory. Finally, we present the results of the measurements of the radon content of various ground and underground water sources.
\end{abstract}

\end{frontmatter}

\section{Introduction}
\noindent

The detection and elimination of the volatile component of the radioactive background that is produced by $^{222}$Rn and its decay products is an essential issue for low background experiments such as dark matter and double beta decay searches \cite{Novella2018,Darwin2016,Albert2018,LUX-ZEPLIN2020NIMA}.
Radon is a noble gas, and therefore it easily penetrates into the shielding of the installation and can undergo decay near the detector, increasing the number of background events.
When the shielding materials or detector components have been in an environment with a high $^{222}$Rn content for a long time, an excess (compared to the equilibrium) number of $^{210}$Pb atoms (T$_{1/2} = 22.3$ years) can accumulate in the surface layer. 

The recoil nuclei $^{210}$Pb arising from the alpha decay of $^{214}$Po can be implanted into the surface of the materials used.
The decay of $^{210}$Pb and daughter isotopes will give an additional background of electrons and alpha particles \cite{Gavr09PTE,Bunker2020NIM,Abe18NIM}.
Consequently, at all stages of the installation, such as during storage, transportation of detectors and, finally, during long-term measurements, it is necessary to know the concentration of radon in the air and take measures to reduce it (see \cite{LUX-ZEPLIN2020EPJC,Aprile2021M,Pelczar2021,Battat2014,Bonet2021}).

In the atmosphere of closed laboratory or underground rock excavations without the use of appropriate protective measures, the concentration of radon can reach high values (up to 1,000 Bq/m$^3$ and higher). Through forced ventilation, radon concentration can be significantly reduced. Nevertheless, the detection and elimination of the volatile components of the radioactive background generated by $^{222}$Rn and its decay products is an urgent problem for ultra-low-background experiments.
$^{222}$Rn is an intermediate volatile radioactive isotope of the radioactive $^{238}$U series.
As a consequence, a radon activity detector is usually the central element of a system that monitors background radioactivity in the air.

An optimal detector seems to be the one that utilizes air as a working medium.
It should have a large enough volume to provide high sensitivity and good energy resolution in order to separate the $\alpha$-particle peaks from those due to decays of $^{222}$Rn and its daughter $\alpha$-active isotopes (e.g. $^{218}$Po, $^{214}$Po, etc.) that are also contained in the sample.

A multi-wire high-sensitivity pulsed air ionization chamber with a large working volume (16 L) has been developed in the Baksan Neutrino Observatory (BNO) of the INR RAS \cite{Kuzminov2003}. Under the conditions of reduced acoustic and vibrational noise, this chamber achieves the energy resolution of 3.9\% for $\alpha$-particles with the energy of 5.49 MeV, produced in decays of $^{222}$Rn. Based on this chamber, we developed a radon monitor and carried out the first long-term measurement of $^{222}$Rn content in the air of both ground, and underground laboratories of the BNO \cite{Gavr11B}.

We found the average radon activity to be $\sim6$ Bq$\cdot$m$^{-3}$ in the open space, $\sim35$ Bq$\cdot$m$^{-3}$ in the air of the ground laboratory (during the measurement that was carried out over the period from February to March of 2004), and $\sim29$ Bq$\cdot$m$^{-3}$ in the air of the KAPRIZ underground laboratory (during the measurement that was carried out over the period from October of 2004 to April of 2005). The underground activity of radon was observed to increase with the air temperature rise, and the average activity  reached the value of $\sim42$ Bq$\cdot$m$^{-3}$ in April of 2005.

The multi-wire chamber had a high sensitivity to microphone noise. As a result, those measurements that had been performed in the underground laboratories equipped with a resounding electric installation were of poor quality and, therefore, were suspended.
In order to overcome the ascertained shortcomings, we developed a cylindrical ion-pulse ionization air chamber (CIPIC) with a reduced microphone noise \cite{Gavr15NIM}. In laboratory conditions, the CIPIC achieves the energy resolution of 1.7\% for $\alpha$-particles with the energy of 5.49 MeV.
A series of $^{222}$Rn activity measurements were carried out with the help of this chamber in various laboratories of the BNO INR RAS. The results of these measurements are presented in this paper.

\section{EXPERIMENTAL METHODS}

\begin{figure}[!th]
\begin{center}
\includegraphics[width=10cm]{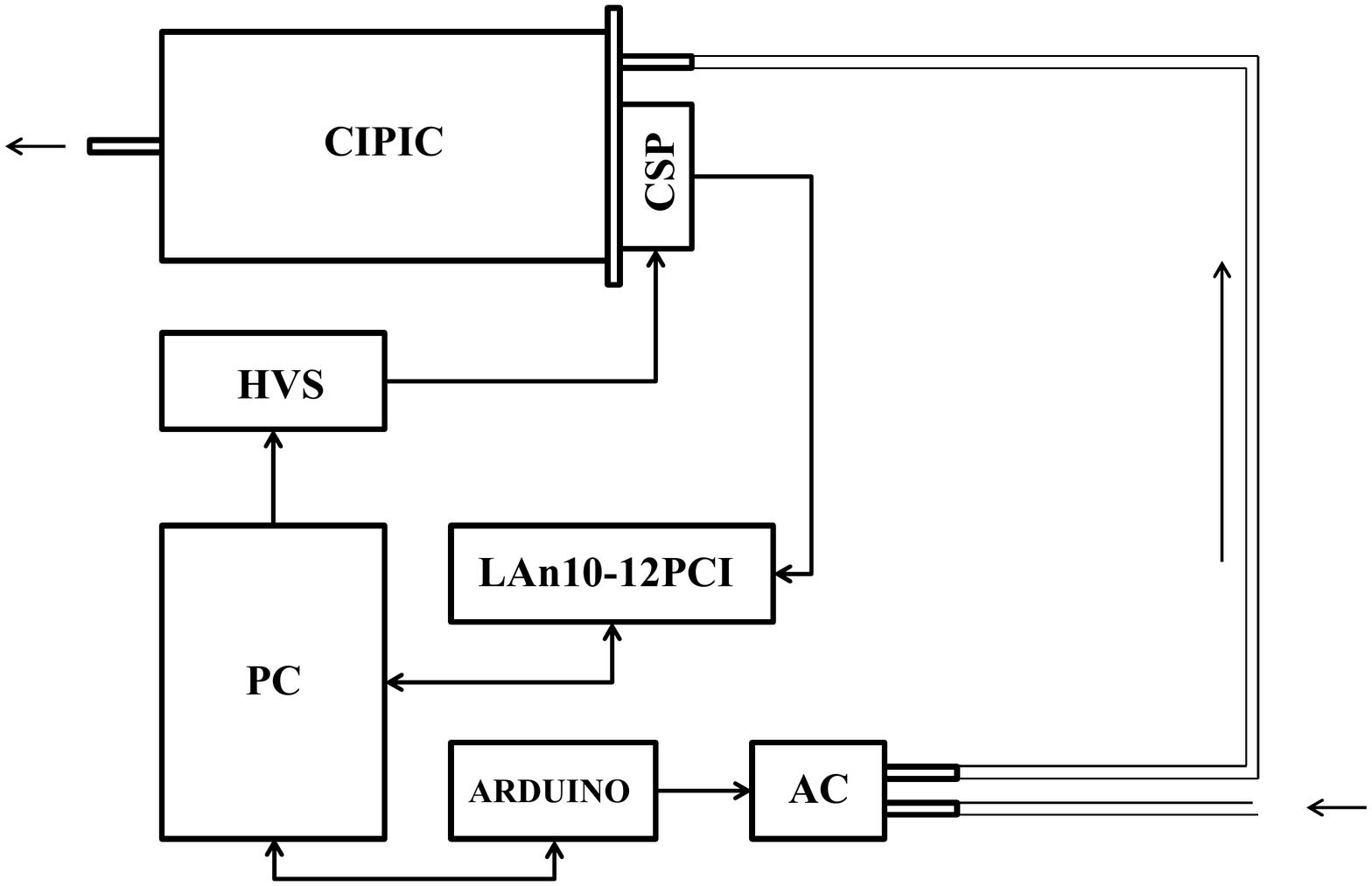}\\
\caption{Block diagram of the system that monitors radon content in the air. The figure contains the following elements: the cylindrical ion-pulse ionization air chamber (CIPIC), the charge sensitive preamplifier (CSP), the high voltage supply (HVS), the personal computer (PC), the digital oscilloscope LA-n10-12PCI, the air compressor (AC) and the ARDUINO controller card.}
\label{Bl-sx-CIPIC}
\end{center}
\end{figure}

In Fig.~\ref{Bl-sx-CIPIC}, we display a block diagram of the radon content monitoring system designed based on the CIPIC.
The working volume $(Vc)$ of the CIPIC is 3220 cm$^{3}$. The air compressor (AC) pumps an air sample from the observation point into the CIPIC. The air is pumped along the pipeline through the Petrjanov filter and silica gel dryer. In order to fully replace the old air sample, more than 10 L of the new air sample is pumped through the CIPIC. The amount of air is adjusted depending on the productivity of the AC and its operation time. After the purge, the chamber relaxes and then the measurement mode is switched on. The high voltage supply (HVS) applies the service voltage on the CIPIC, and that voltage is maintained during the purge. The charge sensitive preamplifier (CSP) reads off the pulses from the CIPIC. The radon monitor operating mode is controlled by the programmable board ARDUINO. The typical measurement time is 6900 s, that of the purge is 240 s, and the typical relaxation time of the CIPIC is 60 s. The blocking relay switches the AC on and off. After that, the cycle repeats. The productivity of the AC is equal to $\sim3.0$ L$\cdot$min$^{-1}$. The measurement and purge times can vary depending on the particular measurement conditions. The digital oscilloscope LA-n10-12PCI records pulses with a sampling frequency of 1.56 MHz. We also used two identical radon monitors.

\section{Results of the measurements}

\subsection{Radon content monitoring in the air of the above ground laboratory}

The data analysis was carried out in the ``offline'' mode. For each measurement cycle, we studied the energy spectrum of the pulses.
When processing the pulses, we accounted for the finite discharge time of the CIPIC's charge-sensitive preamplifier. A detailed description of processing the pulse shape to restore the value of the total charge of the ionization in CIPIC is given in Ref.~\cite{Gavr15NIM}. As a result, we could eliminate the spectrum smearing that stemmed from different pulse rise times.

\begin{figure}[!th]
\begin{center}
\includegraphics[width=12cm]{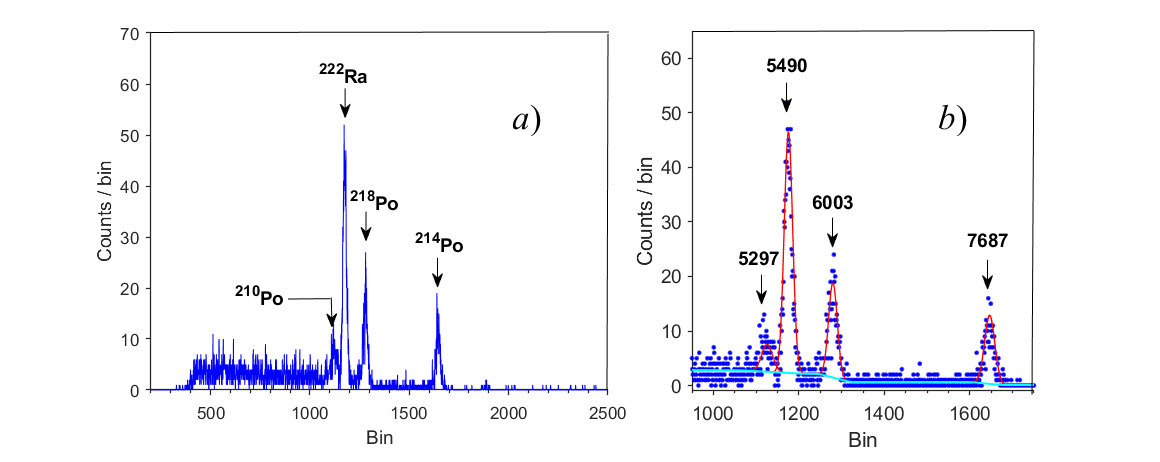} 
\caption{An example of the reconstructed amplitude spectrum of CIPIC pulses, accumulated over the time of 175 minutes in the air of the above-ground laboratory during the measurement of radon content: \emph{a}) - total height spectrum and \emph{b}) - the fitting result. The red curve shows fitting the peaks with a set of Gaussians after subtracting the flat background (birch line). The numbers indicate the ennergy of the $\alpha$-peaks in keV.}
\label{fig02}
\end{center}
\end{figure}

In Fig.~\ref{fig02}\emph{a}, we display an example of the reconstructed spectrum that was collected over the period of time of 175 min. The measurement was carried out on the second floor above the ground laboratory. $\alpha$-lines with the energies of 5.297 MeV ($^{210}$Po), $\sim$5.490 MeV ($^{222}$Rn), 6.003 MeV ($^{218}$Po), 7.687 MeV ($^{214}$Po) can be seen in the spectrum. The area under the peaks was determined by fitting with sets of Gaussians after subtracting the flat background (see Fig.~\ref{fig02}\emph{b}).
The radon content in the air was determined from the area under the Rn-peak in the spectrum during long series of measurements.

\begin{figure}[!th]
\vspace{-7.0pc}
\begin{center}
\includegraphics[width=8cm]{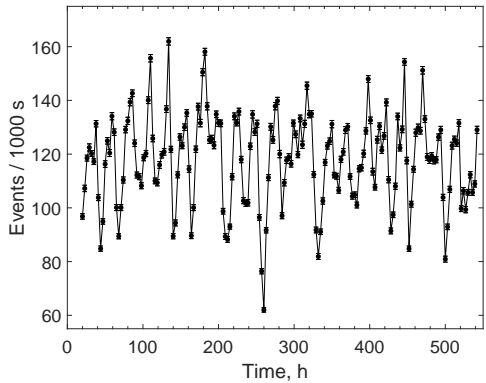}
\vspace{-5.0pc}
\caption{$^{222}$Rn count rate variations in the air of the above ground laboratory over the period from August 25, 2020 to September 16, 2020.}
\label{Rn_count_rate}
\end{center}
\end{figure}

The count rate variability in the air of the above-ground laboratory is shown in
Fig.~\ref{Rn_count_rate}.
The measurement was carried out from August 25, 2020, to September 16, 2020. Daily variations of radon activity related to ventilation during regular working hours and possibly related to the daily atmospheric and tidal effects are well visible. The average count rate of the measurement is equal to $0.117\pm0.016$ s$^{-1}$.

The detection efficiency $(\varepsilon)$ for the 5.49 MeV $\alpha$-particles produced in decays of $^{222}$Rn, which is uniformly distributed in the air volume inside the chamber, is $\varepsilon=0.48$ \cite{Gavr15NIM} at the pressure of 620 Torr. The latter is the air pressure at the height of 1700 m above sea level at the location of the BNO INR RAS. The actual volume activity of radon ($A$) in the air can be calculated as follows,
\begin{equation}\label{a1}
  A=(\varepsilon V_c)^{-1}\times S_R=0.647\times S_R {~}[{\rm Bq\cdot m^{-3}}],
\end{equation}
where $S_R$ is a count rate per 1000 s under the radon peak. This relation takes into account the volume of the chamber and the detection efficiency. This relation takes into account the volume of the chamber and the detection efficiency. Thus, the corresponding activity of $^{222}$Rn in the laboratory was found to be 75.7 Bq$\cdot$m$^{-3}$. As mentioned, a similar measurement in this laboratory was performed with the help of the multiwire chamber over the period from February 26, 2004, to March 9, 2004. The average volumetric activity of radon was $\sim35$ Bq$\cdot$m$^{-3}$.
The last result is consistent with the old one when one takes into account the seasonal changes of radon content in the air.

\subsection{Radon content monitoring in the air of the underground hermetically sealed well}

The surface of the underground tunnel ``MAIN'' and other operational underground laboratories are covered with a $\sim40$ cm thick layer of concrete. In contrast, some of the non-operational premises, reserved excavations, and a considerable part of the tunnel ``AUXILIARY'' are not covered with concrete.
In some places of the underground complex, sources of gas and water have arisen during its existence.
These springs can be the sources of the radon if there is exists a significant amount of the parent $^{238}$U in a surrounding rock and the component material have a substantial enough coefficient of the fluid emanation through the cavities.

\begin{figure}[!th]
\vspace{-4.0pc}
\begin{center}
\includegraphics[width=8cm]{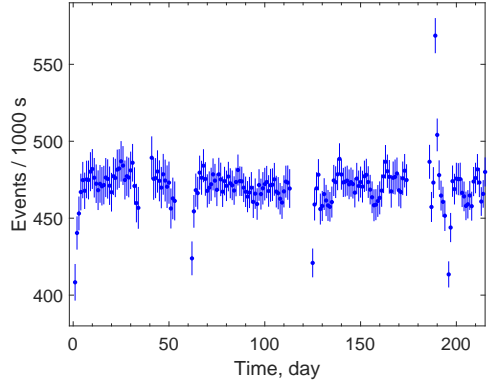}
\vspace{-5.0pc}
\caption{$^{222}$Rn count rate variations in the air of the control hole over the period of time from  January 24, 2020 to July 27, 2020.}
\label{Rn_control_hole}
\end{center}
\end{figure}
Special measurements were done to estimate the value of the possible radon emission from the rock. Measuring set-up was placed in a hall of the DULB-4900 underground low background laboratory \cite{Gavr13NIM}.
A horizontal control borehole (pit) with a diameter of 10 cm and length of 450 cm [volume $\sim35.3$ l] was holed in the rocky wall of the ``Far Geophysical Laboratory'' (FGL) located in the by-lane of the  tunnel ``AUXILIARY'' at the depth of $\sim4000$ m below the ground level \cite{kuzm12}.
The bottleneck of the borehole was sealed with a hermetic plug. Inlet and outlet tubes with an inner diameter of 12 cm were inserted into the hole through the plug. These tubes are 150 m in length and were utilized to connect the well with the radon monitor located inside the laboratory DULB-4900. The tubes' volume $(V_t)$ is $\sim35.9$ L. Due to unfavourable operational conditions of the equipment inside the FGL, we had to separate the monitor
and the control well. The measurements were carried out from January 24, 2020, to July 27, 2020. The count rate variations graph is shown in Fig.~\ref{Rn_control_hole}. We performed a series of methodical corrections to improve the quality of the measurements. The increase of the count rate observed at the beginning of each measurement is tied to the time required for the formation of radon equilibrium inside the measurement volume.

Under the steady-state conditions corresponding to the time interval between the 13th and 33rd day of observations, the area of the Gaussian (the response of the radon signal) is $S_R=(474\pm12)$ counts per 1000 s. The intrinsic background of the closed-loop system (excluding the control hole) had been measured prior to the measurement.
The corresponding count rate of intrinsic background $(S_B)$ was found to be $S_B=(34\pm1)$ counts per 1000 s. Using the values above, we calculated the specific activity of radon in the working air of the chamber $(A_s)$ due to the radon emanation from the walls of the well
\begin{equation}\label{a2}
   A_s=\varepsilon V_c(S_R-S_B)=0.285\pm0.008 {~}[s^{-1}L^{-1}].
\end{equation}
In order to determine the rate of the radon release from the walls of the well, the resulting specific activity should be recalculated, taking into account the activity of the total working volume of the device and the control well. We assume that the balance between the radon release and its decay has been achieved.

This specific activity should be recalculated for the overall volume of the set-up. It is assumed that there is an equilibrium between a radon output and its decay. The calculation accounts for the sealing of the first 20 cm of the well with compaction.  Then the volume of the hole $(V_{c.h.})$ and its surface area $(Q_{c.h.})$ are $V_{c.h.}=33.8$ L and $Q_{c.h.}=13587$ cm$^2$.
The total activity of radon produced in the hole  $(A_{c.h.})$ is
\[
A_{c.h.}=A_s(V_{c.h.}+V_t+V_c)=(18.1\pm0.5){~} {\rm s^{-1}}.
\]
As a result, the specific radon emanation rate ($s$) of the rock is found to be $s=A_{c.h.}/Q_{c.h.}= (13.3\pm0.4)$ s$^{-1}$m$^{-2}$.
One should note that our estimation does not account for the possible influence of the wall roughness on the radon emanation rate.

\subsection{Radon content monitoring in the air flow along the  tunnel MAIN}

\begin{figure}[!th]
\begin{center}
\includegraphics[width=9cm]{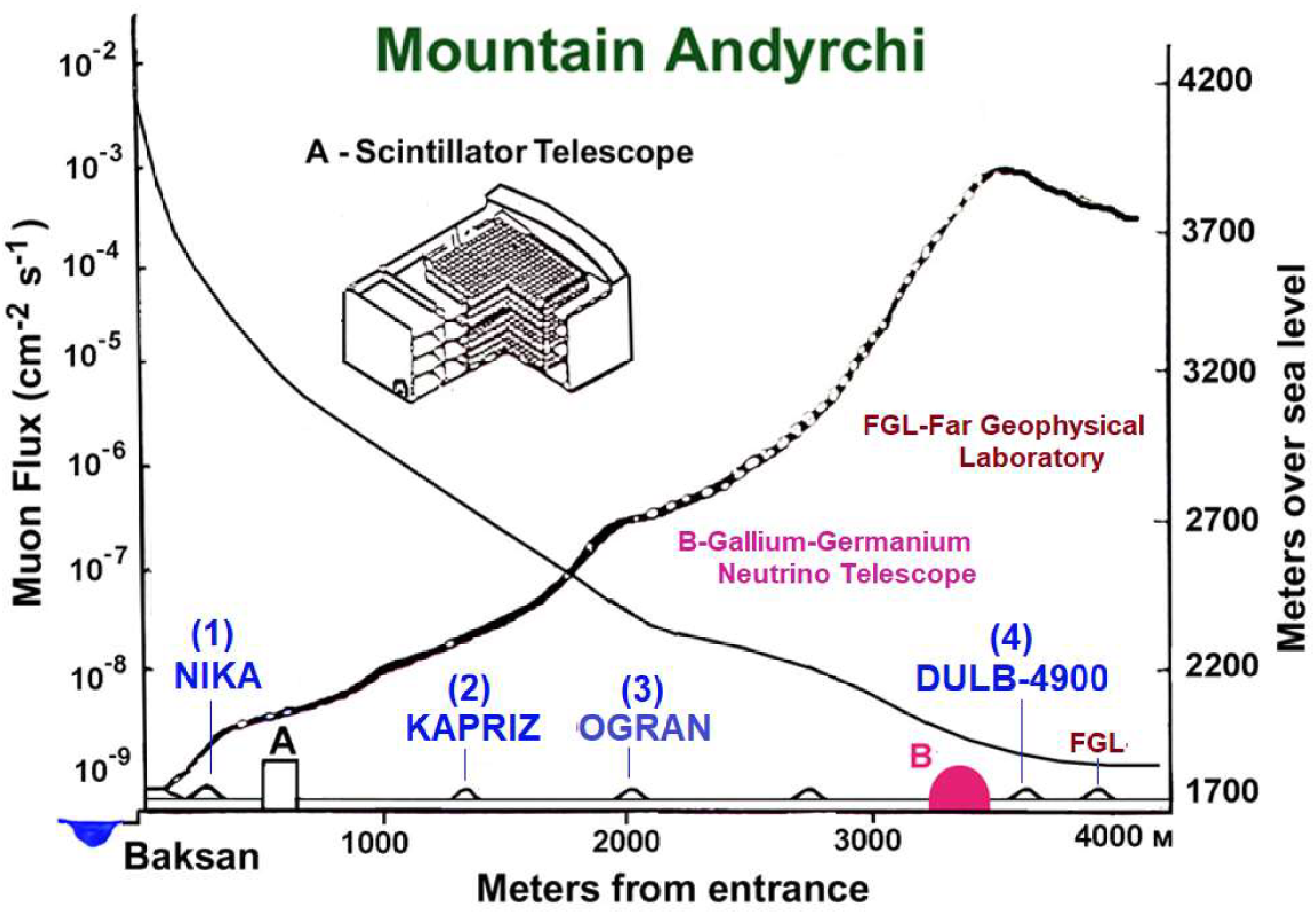}
\caption{Schematic view of the Andyrchi slope section along the adit (right scale) and dependence of underground muon flux on the laboratory location depth (left scale).}
\label{Andyrchi}
\end{center}
\end{figure}

\begin{figure}[!th]
    \centering
    \vspace{-6pc}
    \includegraphics[width=10cm]{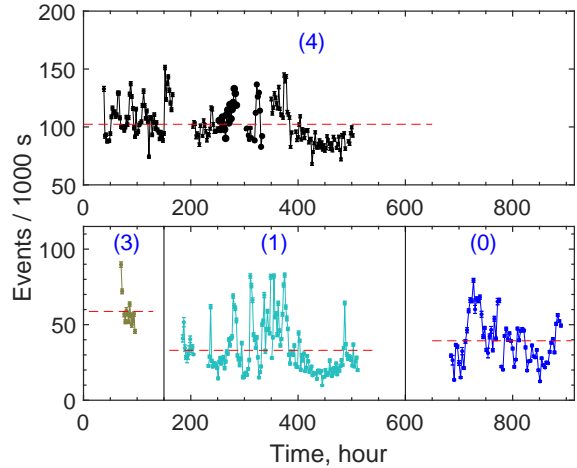} \vspace{-7.0pc}
\caption{$^{222}$Rn count rate variations in the MAIN tunnel air flow: (0 m) - external air; (1) - ``NIKA'' (400 m distance from the entrance)], (3) - ``OGRAN'' (1420 m) and (4) - ``DULB 4900'' (3700 m). The upper dependence was measured with a monitor M1 and the low one was measured with the monitor M2.}
    \label{plt_fig7_M1_2}
\end{figure}

The radon content monitoring in the airflow along the tunnel MAIN (see Fig.~\ref{Andyrchi}) was performed over the time period from January 19, 2021 to February 25, 2021. The two monitors were placed inside different stationary underground laboratories.
The air inside the tunnel was driven by the ground ventilator, which  exhausted the air from the tunnel "AUXILIARY". Four laboratories were used for the measurements: (1) - ``NIKA'' (400 m below the ground level), (2) - ``KAPRIZ'' (620 m), (3) - ``OGRAN'' (1420 m) and (4) -``DULB 4900'' (3700 m) \cite{kuzm12}.
When measuring the radon content in the outdoor air (0 m), the monitor was located on the second floor of the Laboratory Building (LB). The air intake was carried out outside the building via a tube.
The obtained count rate per $10^3$ s is shown in Fig.~\ref{plt_fig7_M1_2}.
Measurements in different locations of the tunnel were carried out over different time intervals. The upper dependence on Fig.~\ref{plt_fig7_M1_2} was obtained using the M1 monitor, and the lower one - using the M2 monitor. The vertical solid lines on the lower graph of Fig.~\ref{plt_fig7_M1_2} segment the plot by laboratory's location. The horizontal dashed lines represent the average count rate values under the 5.49 MeV peak for a given laboratory.
As it an be seen from the graphs, the  radon content experiences large variations of the amplitude at all control points. The monitor M1 tested the air at the location (1) of the tunnel MAIN and the monitor M2 - at the location (4).
It is seen from the graph that the radon content variations at different depths correlate with each other. However, the average values differ approximately by a factor of $\sim3$.

\begin{figure}[!ht]
\vspace{-7.15pc}
    \centering
    \includegraphics[width=8cm]{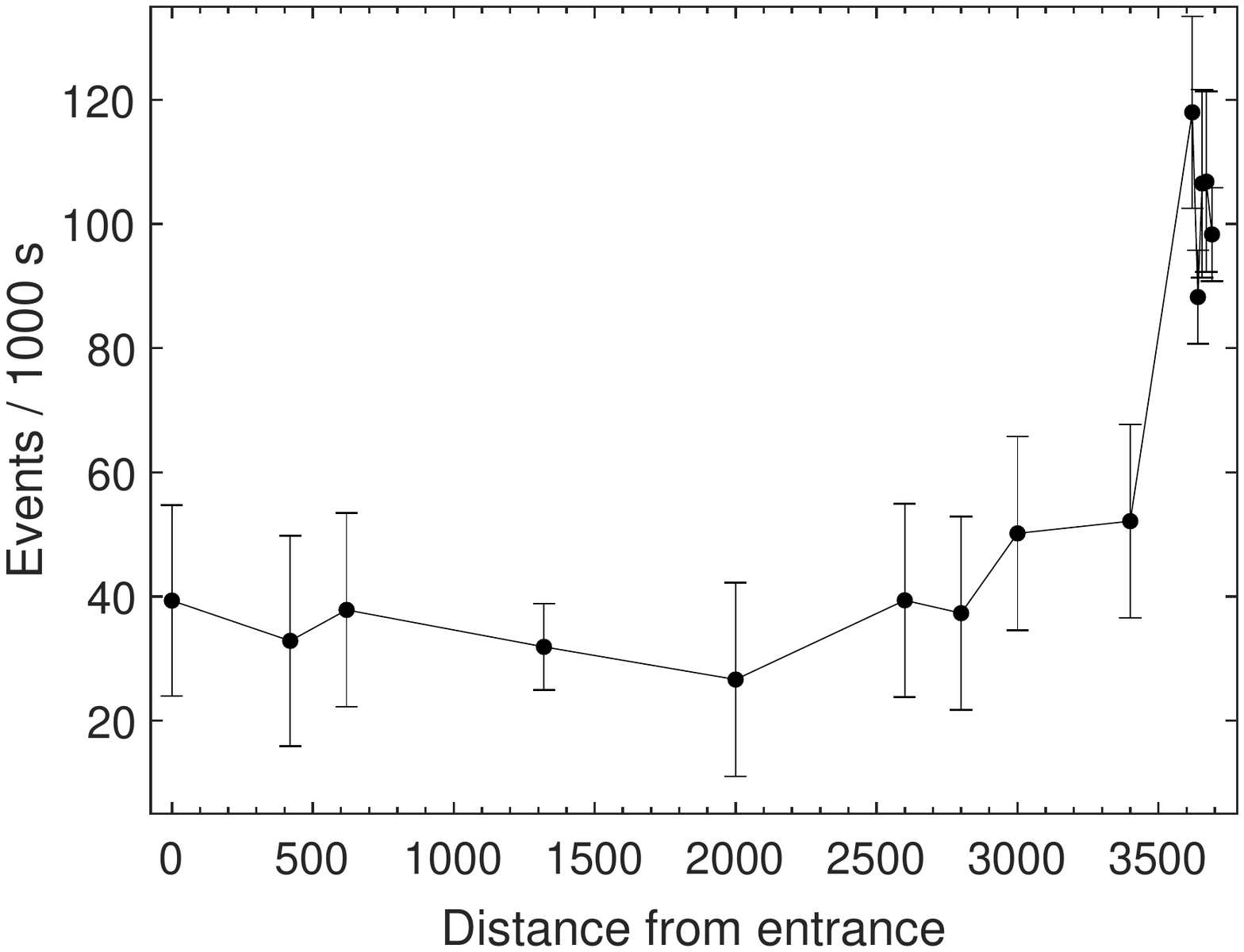} \vspace{-5.85pc}
    \caption{$^{222}$Rn air content variations along the MAIN tunnel.}
    \label{MAIN-tunnel}
\vspace{-7.0pc}
    \centering
    \includegraphics[width=9cm]{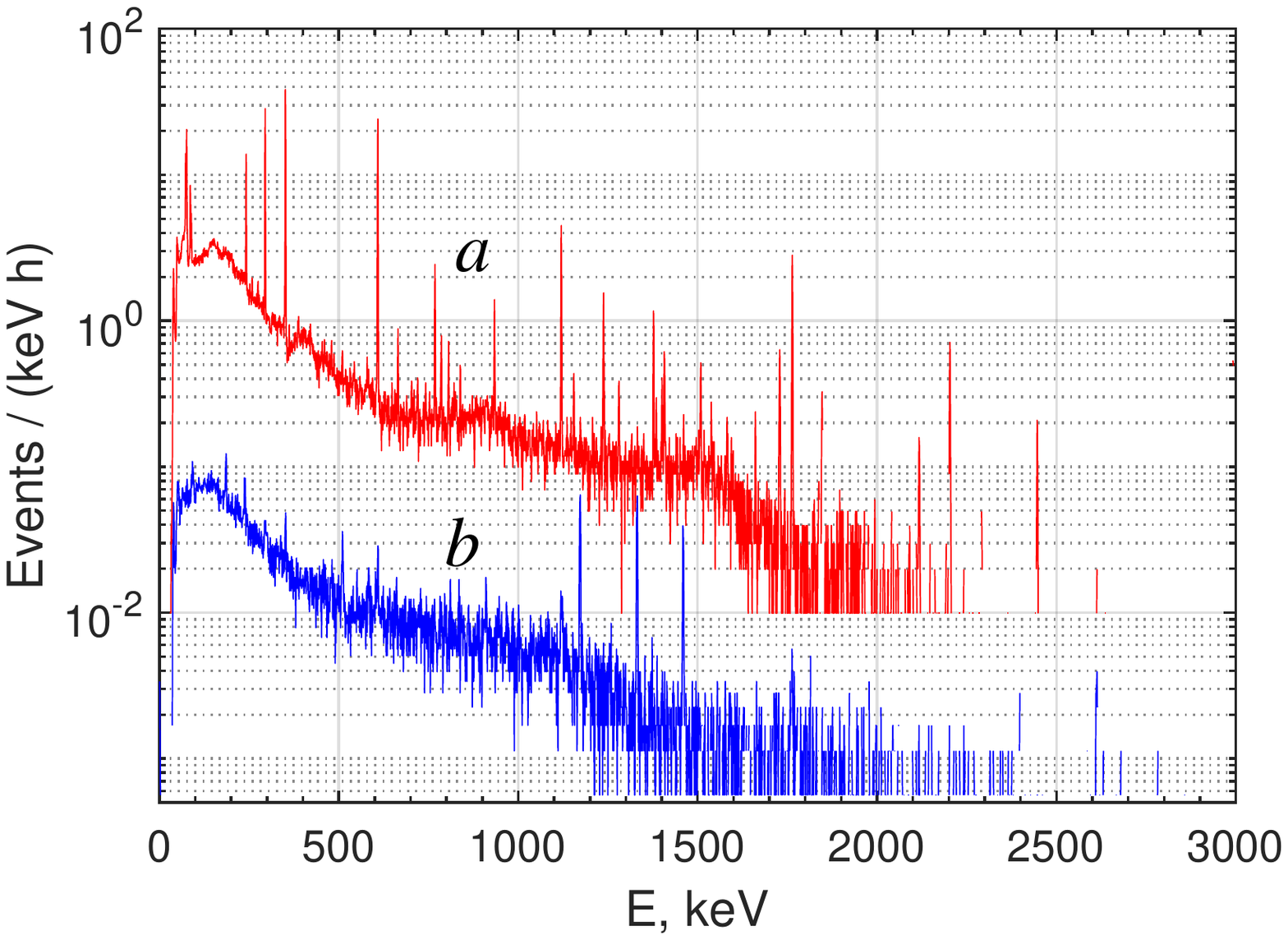}  \vspace{-8.0pc}
    \caption{$(a)$ - amplitude spectrum of radiation from the water sample 1 recorded by the HPGe detector. $(b)$ - background spectrum.}
    \label{spc_HPGe}
\end{figure}

Statistical error of the averaged values does not exceed 5-10\%, but the variation amplitude exceeds the one significantly. Therefore, an uncertainty entered by the variations could be expressed as systematic errors. Averaged values specified in such form depending on the distance between tunnel entrance and a point of measurements are shown in Fig.~\ref{MAIN-tunnel}.

The figure demonstrates that according to the results obtained at stationary points (1) - (3), the radon content in the airflow over a distance of 0-1420 m does not change within the error limits. The radon content increases $\sim3$ times at a distance of 1420-3700 m from the entrance to the tunnel. This behaviour cannot be explained by the contribution to the radon content of gas release from the concrete walls of the adit, which should uniformly increase the content with increasing distance. It was proposed that there were point sources of radon in the tunnel.

The air samples were taken at the depths of 2000 m, 2600 m, 2800 m, 3000 m and 3400 m directly into the volume of the CIPIC, which we used as a sampler. The measurements were performed under optimal operating conditions at locations (3) and (4). Since the measurements were isolated, we used the value of the uncertainty at point (1) as the systematic uncertainty for all measurements. It can be seen from the graph that, even though there are small underground water-gas outlets in the walls of the tunnel at distances of 2600 m and below the ground level, the radon content in the airflow changes marginally down to 3400 m. A rise of the radon content in the air is observed at distances of 3500 m below the ground level, where the main and auxiliary entrances to the Gallium-Germanium Neutrino Telescope laboratory (GGNT) are located \cite{kuzm12}.

There is another possible source of radon. The GGNT uses conditioned air to maintain a proper temperature regime inside the laboratory rooms. A water conditioner is used for the preparation of the conditioned air. The input airflow is transmitted through water drops to filter out dust and then through the water heat exchange to achieve cooling. The total water flow rate is $\sim4$ L$\cdot$s$^{-1}$. The conditioner is fed by potable water with the help of the external water well. The water is discharged to the open channel, which is laid along the walls of the tunnel. The water mixes considerably and is exposed to the air over tens of meters.

\begin{table}[!ht]
    \caption{$^{222}$Rn activity in the water samples from different sources of the BNO INR
    RAS}
    \label{tab:results}
\hspace{1.0pc}
    {\centering
    \begin{tabular}{|r|l|c|c|c|}\toprule
 ~   &      ~         &            ~       &        ~    & Initial $^{222}$Rn activity  \\
 No. &\multicolumn{1}{c|}{Water samples}   & 609 keV & Collect.time& in the water sample \\
 ~   &      ~         &     peake area    & (hour)      &  (Bq$\cdot$kg$^{-1}$)   \\\midrule
1 &Exhaust tube of the GGNT conditioner	           &4520 &	101.2 &	$12.1\pm0.2 $ \\
2 & Channel in front of the GGNT gate$^*$          &1712 &	101.2 &	$4.6 \pm0.1 $ \\
3 & Tap in the ``NIKA'' lab	                           &6369 &	  69  &	$22.3\pm0.3 $ \\
4 &Intermediate storage water tank	                   &9389 &	106   &	$24.5\pm0.3 $ \\
5 & Water intake	                                   & 8637&	 99   &	$23.4\pm0.3 $ \\
6 & Water-gas leak (1) at the GGNT wall               & 530 &	 54   & $2.5 \pm0.1 $ \\
7 &Water-gas leak (2) at the GGNT wall                & 542 &	 51   & $2.4 \pm0.1 $ \\
8 & Birch sap	                                       &24.5 &	139   & $0.04\pm0.01$ \\
9 & Baksan river water                                & 99  & 78.8   & $0.31\pm0.03$ \\
10& Gubasanty stream water                            &125  & 182.8  & $0.20\pm0.02$ \\
11& Roadside spring                                   &5880 & 185.6  & $10.5\pm0.1 $ \\
12& Spring near boiler-house                          &2180 & 166.5  & $4.3 \pm0.1 $ \\
13& Mineral water	                                   &7698 &   69   & $27.1\pm0.3 $ \\
14& Own set-up background                             &  15 & 1776   & $0.021\pm0.005$\\
    \hline
\end{tabular}} \newline{$^*$75 m below No.~1}

\end{table}

Radon that comes out of the water should contribute to the radon activity of the passing air. The amount of the deposited radon depends on the radon content of water and the intensity of the gas exchange.

The radon content of water was determined from the activity of the 609 keV $\gamma$-line of the daughter isotope $^{214}$Bi by using the low-background high-purity germanium (HPGe) detector. In order to exclude those $^{214}$Bi decays that cannot be traced back, the data were recorded 3 hours after the measurement had been started. Two water samples were used for this measurement. The first one was a sample taken from the place of discharge, and the second one was taken 75 m down the stream. The results of this measurement are shown in Table \ref{tab:results}.

The radon activity of the sample (1) and (2) was found to be  $(12.1\pm0.2)$ Bq$\cdot$L$^{-1}$ and $(4.6\pm0.1)$ Bq$\cdot$L$^{-1}$, respectively. The radon inside the downpipe of the conditioner could leak into the air because water occupies only a part of the pipe's cross-section. Additional measurements were carried out in order to determine the overall content of radon in water and understand the dynamics of radon's behaviour inside the pipe. The corresponding measurements were performed using the water probes taken from the tap at NIKA [sample (3)], the intermediate water tank (4), and the water intake (5). It can be seen from a comparison of the data that the radon content inside the water moving along the airproof water path at the pressure of $\sim3$ bar does not change much. Therefore, the radon content inside the water delivered to the conditioner input should be the same as inside the sample (3). In this case, it is necessary to take into account that some amount of radon from the sample (3) could leak out of the water together with micro bulbs of air formed after the pressure dump had occurred during the sample splitting.
\begin{figure}[!ht]
\hspace{-5.0pc}
\vspace{-11.0pc}
\begin{center}
    \includegraphics[width=12cm]{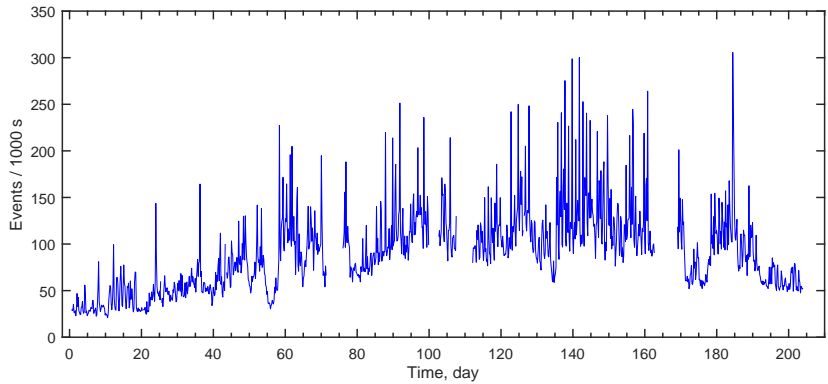}\vspace{-13.0pc}
    \caption{Time dependence of the air $^{222}$Rn content in the point (2) of the MAIN tunnel. The measurements were carried out from March 16, 2021, to October 06, 2021.}
    \label{dependence_air}
\end{center}
\end{figure}

The spectrum $(a)$ of the $\gamma$-radiation from the water sample (1) is shown in Fig.~\ref{spc_HPGe}. This figure illustrates the capabilities of a facility with the HPGe detector. The volume of this and other samples is 170 cm$^3$. The background spectrum $(b)$ is also shown here.
The water samples (6) and (7) from the water-gas leaks at the walls of the GGNT were taken in order to estimate the corresponding radon deposits. Due to a low flow rate of water,  such leaks could give a relatively small deposit to the radon activity.
The results of the measurement are shown in Table \ref{tab:results}.

One should expect that the volume activity of radon could increase up to $\sim50$ Bq$\cdot$m$^{-3}$, in dependence of a passing air rate when radon is entirely released from the water used by the conditioner. Apparently, this source is the main reason for the radon activity rise at the 3600 m location at the MAIN tunnel.

\subsection{Seasonal variations of the radon content in the air of the tunnel}

As it can be seen from Fig.~\ref{MAIN-tunnel}, the radon content in the flowing air of the tunnel does not differ much from that in the open space.
Therefore, in order to observe the seasonal variation of radon in the external air, we employed the data from the monitor located at the point (2). The corresponding time dependence of the measurement over the period from March 16, 2021 to October 6, 2021 is shown in Fig.~\ref{dependence_air}.
It can be seen that the average content of radon in the air increased by $\sim4$ times by the end of June.
The reason for this growth may be the following seasonal process. At the beginning of the measurements, the area around the BNO INR RAS was covered with snow. The air temperature ranged from $-10$ $^\circ$C to $0$  $^\circ$C. Since March 26, the daytime temperature has shifted to the region of positive values, and by the beginning of May, the slopes of the hills were cleared of snow to a height of $\sim2500$ m.

In the surrounding area, the trees and grass turned green. The corresponding processes are accompanied by the increase of water transfer from the soil to the plants in the form of a plant sap. If that water contained radon, it would have been released to the air from the plants through a leafy surface. In order to check this assumption, we measured the content of the daughter $^{214}$Bi  in a birch sap with the help of the HPGe detector. The results of our measurement for the sample 8 are presented in Table \ref{tab:results}. As it is seen from the table, the content of radon in the sample was minor.

The water flow of the Baksan river increases by multiple times during the summer period. If radon was contained in that water, the intensive mixing of water streams would have been released into the air. Therefore, we also measured radon content in the sample of river water with the help of the HPGe detector. The results of the measurement for the sample 9 are presented in Table \ref{tab:results}.
The content of radon in the sample is minor. Besides the river, there are several streams and springs in the vicinity of the BNO INR RAS. The results of radon content measurements in some of the corresponding samples are shown in Table \ref{tab:results}, too. It can be seen that the largest content of radon is associated with the samples of spring water. The water of surface streams consists mainly of the glacial meltwater. The glacier ice contains a low concentration of mother isotopes of radon, and thus, the radon content in glacial melt water is low, too. During the winter season, the water of springs with a high radon content gives a significant rise to the radon activity of the river in the locations of the water inflow.
Following the data for the samples 1 and 2 provided in Table \ref{tab:results}, on can conclude that the radon content decreases with the increasing distance from the inflow of spring water.
Moreover, one can see that passing the first hundred meters from the inflow of spring water, radon entirely escapes the river.

When the land surface dries out, soil capillaries open up during the warm weather. It seems that the radon released from the soil through capillaries is the primary source of radon in the open air. Therefore, prolonged, extensive rains should wash radon out from the atmosphere and occlude the capillaries. As a result, the radon content in the air should decrease. A substantial short-time emission of radon could be caused by wind gusts, which forcefully extract radon from the earth stratum in a way similar to how a pump operates. The resulting local rise in the above-ground content of radon is dissipated quickly due to its mixing with relatively clean masses of the off-site air. The decrease of the radon content could occur during periods of strong winds, when relatively clean high-altitude air layers mix with the above-ground air, provided that the radon content in it does not exceed its locally made values. This issue requires further study.

\section{Summary and conclusions}

Our measurements performed with the help of the radon monitor device suggest that it has good performance characteristics and retains them during long (several months) measurements.
The device was developed based on the cylindrical air ion pulse ionization chamber. At moderate humidity levels, the chamber operates stably without any preliminary air swabbing, as it was observed, for instance, during the winter period. The detector has a reduced level of microphone noise. It allows achieving a high energy resolution at the level of 1.5-2.0 \% when registering 5.49 MeV energy releases from decays of $^{222}$Rn and its daughters. One of the two radon monitors was used to determine the release of radon from the surface of the rocky ground.  It was determined that the rate of the radon emanation from the inner surface of a cylindrical hole, which is 10 cm in diameter and 450 cm in length, was $(13.3\pm0.4)$ c$^{-1}\cdot$m$^{-2}$. Continuous measurements have shown that variations of the rate do not exceed $\sim2$\% over the period of time of $\sim200$ days.

A concurrent operation of two monitors allows one to measure the radon content simultaneously in separate locations of the tunnel. Our results show that, within the precision of the measurement, the radon emanation from the concrete walls of the tunnel does not provide an additional contribution to the radon content of the flowing underground air. This coincides with our expectations, calculated under the assumption that the emanation rate of  radon from a concrete surface is identical to that from a rocky ground. Down to $\sim3400$ m below the ground level, the radon content in the flowing air of the tunnel has no notable difference from that in the outside air. The corresponding levels of radon content undergo simultaneous fluctuations with the amplitudes which could differ by 1.5-2 times from the average values. The increase of the radon content in the air was observed at distances of 3400 m below the ground level. With the help of the semiconductor detector that measured the radon content of various water samples collected at the territory of GGNT, it was determined that the main source of radon is the potable water used by the air-conditioner which is then discharged to the open groove. The released from this water radon  increases the levels of radon content by more than $\sim2$ times in comparison with the average winter value, which is $\sim30$ Bq$\cdot$m$^{-3}$.

This deposit could be eliminated if the water discharge was realized through a drain pipe that goes outside the tunnel.
Another possibility is to use initially pure water or water purified from radon by any of the available methods.
Our tests of water samples collected from different surface sources showed
that wells and springs contain increased levels of radon content. The tests were performed with the help of the HPGe detector. The radon content in the river and stream water samples was ten times smaller than that in spring water samples. One of the main goals of the carried out investigations, which measured radon content variations in the underground air, was to estimate a possible influence of the radon decay background on the results of underground low-background experiments. Continuous observations of radon activity in the air were made over $\sim7$ months because the corresponding experiments are very durational. It was observed that the radon content in the flowing underground air of the BNO INR RAS depends on the season of the year and can vary from $\sim30$ Bq$\cdot$m$^{-3}$ in wintertime to $\sim100$ Bq$\cdot$m$^{-3}$ in  summertime. If the low-background set-up is not shielded enough, the annual background variation of the sought-for effect can appear in experimental data due to radon and its daughter decays. Such a background can be completely eliminated if a proper experimental set-up is designed.

\section*{Acknowledgement}
We want to thank Oleksandr Koshchii (Johannes Gutenberg-Universit\"{a}t Mainz) for productive discussions as well as for solid support during the final stage of the manuscript.
The work is performed in accordance with the INR RAS research plan.

\end{document}